\documentclass[10pt]{article}
\usepackage{amsmath}
\usepackage[latin1]{inputenc}
\usepackage{amssymb, amscd, amsthm}
\usepackage[all]{xy}
\usepackage[dvips]{graphicx}
\usepackage{verbatim}
\usepackage[perpage,symbol*]{footmisc}
\usepackage{cite}
\usepackage[colorlinks, linkcolor=red,anchorcolor=blue, citecolor=green]{hyperref}
\usepackage{longtable}
\newcommand{\tabincell}[2]{\begin{tabular}{@{}#1@{}}#2\end{tabular}}
\usepackage{multirow}

\newcommand{\F}{\mathbb{F}}
\begin{document}
\title{\bf  Entanglement-assisted Quantum MDS Codes from Generalized Reed-Solomon Codes}
\author{Renjie Jin\footnote{jinrenjie@mails.ccnu.edu.cn (R.~Jin)}, Yiran Cao\footnote{741148312@qq.com(Y.Cao)} and Jinquan Luo \footnote{Corresponding author,luojinquan@mail.ccnu.edu.cn (J.~Luo).
\newline
 The authors are with School of Mathematics and Statistics $\&$ Hubei Key Laboratory of Mathematical Sciences, Central China Normal University, Wuhan, China 430079.}}
\date{}
\maketitle

\medskip

{\bf Abstract} \, In this paper, some new classes of entanglement-assisted quantum MDS codes (EAQ
MDS codes for short) are constructed via generalized Reed-Solomon codes over finite fields of odd characteristic. Among our constructions, there are many EAQMDS codes with new lengths. Moreover, some of these EAQMDS codes have larger minimum distance than known results.\\


\section{Introduction}
\noindent Entanglement-assisted quantum error-correcting codes\;(EAQEC codes for short) make use of preexisting entanglement between sender and receiver to boost the rate of transmission. The concept of
EAQEC code was introduced by Brun et al.$[\ref{B}]$, they showed that the entanglement-assisted quantum codes they described do not require the dual-containing constraint necessary
for standard quantum error correcting codes. Entanglement plays an important role in quantum information processing. It enables the teleportation of quantum states without physically sending quantum systems.
For more than one decade, construction of good quantum codes via classical codes quantum error-correcting codes are crucial to quantum information and quantum computation $(\text{see} [\ref{As},\ref{Ca},\ref{Cal},\ref{Ch},\ref{Ke}]).$

Many researchers  have been devoted to obtaining EAQEC codes via classical liner codes, including negacyclic codes and generalized Reed-Solomon codes\;$(\text{see}[\ref{Fang},\ref{FF}, \ref{LR},\ref{Lu},\ref{QJ2},\ref{QJ4}])$.
It has been showed that EAQEC codes have some advantages over standard stabilizer
codes.
Some of them can be summarized as follows. In $[\ref{Ko},\ref{Lin}]$, some new EAQEC codes with good parameters via cyclic and constacyclic codes are constructed. In $[\ref{Che}]$, a new decomposition of negacyclic codes is proposed, by which four new classes of EAQEC codes have
been constructed. In $[\ref{Fan}]$, Fan et al. constructed some classes of EAQMDS codes based on classical MDS codes by exploiting one or more pre-shared maximally
entangled states. In $[\ref{QJ3}]$, Qian and Zhang constructed some new classes of MDS linear complementary dual(LCD) codes with respect to Hermitian inner product. As applications, they constructed new families of EAQMDS codes. In $[\ref{Gu}]$, Guenda et al. showed that the number of shared pairs required to construct an EAQEC code is related to the hull of classical codes. Using this fact, they put forward new methods to construct EAQEC codes with desired parameters. In this paper, we will construct EAQMDS codes via generalized Reed-Solomon code.\\

 We will present some known results of EAQMDS codes, which are depicted in Table 1. Suppose $q$ is an odd prime power.
\begin{center}
\begin{longtable}{|l|l|l|}  
\caption{Some known  EAQMDS codes with parameters $[[n,k,d;c]]_q$}\\ \hline
Parameters & Constraints & Reference\\ \hline
$[[q^{2}+1,q^{2}-4(m-1)(q-m-1),$
$2(m-1)q+2;4(m-1)^{2}+1]]_{q}$
  &  $q\geq 5$ and $2\leq m\leq\frac{q-1}{2}$   & [\ref{QJ4}] \\  \hline

$[[q^{2}+1,q^{2}-2q-4m+5,$
$2m+q+1;4]]_{q}$ & \tabincell{c}{$q\geq 5$, $q=4t+1$ where $t$ \\is an integer and $2\leq m\leq\frac{q-1}{2}$} & [\ref{Che}]\\  \hline

$[[\frac{q^{2}+1}{2},\frac{q^{2}+1}{2}-2q-4m+5,$
$2m+q+1;5]]_{q}$ & $q>7$ and $2\leq m\leq\frac{q-1}{2}$ &   [\ref{Che}] \\  \hline

$[[\frac{q^{2}-1}{5},\frac{q^{2}-5q-20m+4}{5},$
$\frac{4m+q+5}{2};4]]_{q}$ & \tabincell{c}{$q=20t+3$ or $q=20t+7$ \\and $t\leq m \leq\frac{q-3}{4}$} &   [\ref{C1}]\\  \hline

$[[\frac{q^{2}+1}{10},\frac{q^{2}+1}{10}-2d+3,$
$d;1]]_{q}$ & \tabincell{c}{$q=10m+3$ and $d$ is even,\\ $2\leq d \leq 6m+2$} &  [\ref{L}] \\  \hline

$[[\frac{q^{2}+1}{10},\frac{q^{2}+1}{10}-2d+3,$
$d;1]]_{q}$ & \tabincell{c}{$q=10m+7$ and $d$ is even,\\ $2\leq d \leq 6m+4$} &   [\ref{L}] \\  \hline

$[[\frac{q^{2}-1}{h},\frac{q^{2}-1}{h}-2d+3,$
$d;1]]_{q}$ & \tabincell{c}{$h\in \{3,5,7\}$ is a factor\\ of $q+1$ and $d$ is even,\\ $\frac{q+1}{h}\leq d\leq\frac{(q+1)(h+3)}{2h}-1$}  &   [\ref{L}] \\  \hline

$[[\frac{q^{2}-1}{t},\frac{q^{2}-1}{t}-4qm+4m^2+3,$
$2m(q-1);(2m-1)^{2}]]_{q}$ & \tabincell{c}{$q\geq 3$, $t\mid q^2-1$,\\ $1\leq m\leq\lfloor\frac{q+1}{4t}\rfloor$} &  [\ref{J}] \\  \hline

\tabincell{c}{$[[\frac{q^{2}+1}{t},\frac{q^{2}+1}{t}-4qm+4q+4m^2-8m+3,$\\
$2q(m-1)+2;4(m-1)^2+1]]_{q}$}  & \tabincell{c}{$q\geq 7$, $t\mid q^2+1$,\\ $2\leq m\leq\lfloor\frac{q+1}{4t}\rfloor$} & [\ref{J}] \\  \hline

\tabincell{c}{$[[lh+mr,lh+mr-2d+c,d+1;c]]_{q}$}  & \tabincell{c}{$ s\mid q+1,\;t \mid q-1,$\\
$ l=\frac{q^2-1}{s}, m=\frac{q^2-1}{t},$\\
$ 1\leq h\leq \lfloor\frac{s}{2}\rfloor, 2\leq r\leq \lfloor\frac{t}{2}\rfloor,$\\
$ c=h-1 \;\text{and}\; $\\
$1\leq d \leq \text{min}$\\
$\{\frac{s+h}{2}\cdot\frac{q+1}{s}-2, \frac{q+1}{2}+\frac{q-1}{t}-1\}.$} & Theorem 3.1 \\  \hline

\tabincell{c}{$\left[\left[1+(2e+1)\frac{q^2-1}{2s+1},1+(2e+1)\frac{q^2-1}{2s+1}-2k+c,k+1;c\right]\right]_{q}$}&
\tabincell{c}{$0\leq e\leq s-1,\;$
\\$ (2s+1)\mid q+1, c=2e$ and\\
  $1\leq k \leq(s+1+e)\frac{q+1}{2s+1}-1.$} & Theorem 4.1 \\  \hline

\tabincell{c}{$\left[\left[1+(2e+2)\frac{q^2-1}{2s}, 1+(2e+2)\frac{q^2-1}{2s}-2k+c, k+1; c\right]\right]_{q}$}  &
 \tabincell{c}{$0\leq e\leq s-2,\;$\\
 $  2s \mid q+1,   c=2e+1$ and\\
  $1\leq k \leq (s+1+e)\frac{q+1}{2s}-1. $}  & Theorem 4.1 \\  \hline

\tabincell{c}{$\left[\left[1+(2e+1)\frac{q^2-1}{2s}, 1+(2e+1)\frac{q^2-1}{2s}-2k+c, k+1; c\right]\right]_{q}$}  &
 \tabincell{c}{$0\leq e\leq s-1,\;$\\
 $  2s \mid q+1,   c=2e$ and\\
  $1\leq k \leq (s+e)\frac{q+1}{2s}-2. $}  & Theorem 4.1 \\  \hline

\end{longtable}
 \end{center}

This paper is organized as follows. In section 2, we will introduce some basic knowledge and useful results on generalized Reed-Solomon codes and EAQEC codes. In sections 3 and 4, we will present our main results on the constructions of new EAQMDS codes. In section 5, we will give the proof of Lemmas 3.1 and 4.1. In section 6, we will make a conclusion.\\\\

\section{Preliminaries}
\subsection{Generalized Reed-Solomon Codes}
In this section, we introduce some useful results about Hermitian self-orthogonality and
generalized Reed-Solomon codes.\\

Let $\mathbb{F}_{q}$ be the $\emph{finite field}$ of cardinality $q$, where $q$ is an odd prime power. An $[n,k,d]_{q}$ code $\mathcal{C}$ is an
$\mathbb{F}_{q}$-linear subspace of $\mathbb{F}^{n}_{q}$ with dimension $k$ and minimum distance $d.$
$\emph{The Singleton bound}$ states that
$$ d\leq n-k+1.$$
A code attaining the Singleton bound is called  $\emph{Maximum Distance Separable}$ (MDS for short) code. Let $\mathbf{v}\in\F_{q^2}^{n}$ and $\mathbf{v}^{q}=(v_{0}^{q},\cdots,v_{n-1}^{q}).$  $\emph{The Hermitian
inner product}$  of  $\mathbf{u}=(u_{0},\cdots,u_{n-1}),
 \mathbf{v}=(v_{0},\cdots,v_{n-1})\in \mathbb{\F}_{q^2}^{n}$ is defined by $$ \langle \mathbf{u},\mathbf{v} \rangle_{H}=u_{0}v_{0}^q+u_{1}v_{1}^q+\cdots+u_{n-1}v_{n-1}^q.$$
$\emph{The Hermitian dual}$ of an $\mathbb{F}_{q^2}$-linear code $\mathcal{C}$ with length $n$ is defined as $$\mathcal{C}^{\perp_{ H}}=\{\mathbf{u}\in \F_{q^2}^{n} \mid \langle\mathbf{ u},\mathbf{v} \rangle_{H}=0 \;\text{for\,all}\,\mathbf{v}\in \mathcal{C}\}.$$
The code $\mathcal{C}$ is $\emph{Hermitian self-orthogonal}$ if $\mathcal{C} \subseteq \mathcal{C}^{\perp_{H}}$, and is  $\emph{Hermitian self-dual}$ if $\mathcal{C}= \mathcal{C}^{\perp_{H}}$.\\

 Choose two $n$-tuples
$\mathbf{v}=(v_{1},v_{2},\ldots,v_{n})$ and $\mathbf{a}=(\alpha_{1},\alpha_{2},\ldots,\alpha_{n})$, where $v_{i}\in\mathbb{F}_{q}^{*}$,
$1\leq i\leq n$ ($v_{i}$ may be not distinct) and $\alpha_{i}$ ($1\leq i\leq n$) are distinct elements in $\mathbb{F}_{q}$. For an integer $k$ with
$1\leq k\leq n$, the $generalized\; Reed$-$Solomon\; code$ ( $\mathbf{GRS}$ code for short) of length $n$ associated with $\mathbf{v}$ and $\mathbf{a}$ is defined as follows:
\begin{equation*}\label{def GRS}
\mathbf{GRS}_{k}(\mathbf{a},\mathbf{v})=\left\{\left(v_{1}f(\alpha_{1}),\ldots,v_{n}f(\alpha_{n})\right):f(x)\in\mathbb{F}_{q^2}[x],\mathrm{deg}(f(x))\leq k-1\right\}.
\end{equation*}
The Hermitian dual code of $\mathbf{GRS}_{k}(\mathbf{a},\mathbf{v})$ is also an $[n,n-k]_{q^2}$ MDS code.\\
A $generator\; matrix$ of $\mathbf{GRS}_{k}(\mathbf{a},\mathbf{v})$ is given by
\begin{equation}\label{1}
   G=\left(
       \begin{array}{cccc}
         v_{1} & v_{2} & \cdots & v_{n} \\
         v_{1}a_{1} & v_{2}a_{2} & \cdots &  v_{n}a_{n} \\
         \vdots & \vdots & \ddots &\vdots \\
         v_{1}a_{1}^{k-1} & v_{2}a_{2}^{k-1} &\cdots & v_{n}a_{n}^{k-1} \\
       \end{array}
     \right).
\end{equation}

\subsection{EAQEC code}

An EAQEC code can be denoted as $[[n,k,d;c]]_{q}$,  which encodes $k$ information qubits into $n$ channel
qubits with the help of $c$ pairs of maximally entangled states and corrects up to $\lfloor\frac{d-1}{2}\rfloor$
errors, where $d$ is the minimum distance of the code.\\

{\bf Lemma 2.1 } $([\ref{C}])$\quad For any $[[n,k,d;c]]_q$ EAQEC code with $d\leq\frac{n+2}{2}$, $\emph{the EA-Singleton bound}$ for EAQEC codes is
$$2(d-1)\leq n-k+c,$$ where $0\leq c\leq n-1.$

An EAQEC code attaining this bound is called an EAQMDS code.  In the following result, Lv et al. showed that EAQEC codes can be constructed from arbitrary classical linear codes over $\F_{q^2}.$\\

Suppose $C^{m\times n}$ is the set of all matrices whose elements belong to the complex domain $C$. Let $\overline{A}$ be the matrix consisting of the conjugate of the elements of $A$. Denote by $A^{\dag}=(\overline{A})^{T}.$ Then $A^{\dag}$ is the conjugate transpose of $A$.\\

{\bf Theorem 2.1 }$([\ref{Lv}])$\quad Let $\mathcal{C}$ be an $[n,k,d]_{q^2}$ classical linear code over  $\F_{q^2}$ with parity check matrix $H$.
There exists an EAQEC code with parameters $[[n,2k-n+c,d;c]]_{q},$ where $c=rank(HH^{\dag})$ and $H^{\dag}$ is the conjugate transpose of
$H$ over $\F_{q^2}.$

\section{First Construction}
In this section, we present new EAQMDS codes from $\mathbf{GRS}$ codes.
If there are no specific statements, the following notations are fixed throughout this section.
\begin{itemize}
  \item Let $q>3,\;s\mid q+1$ and $t\mid q-1$ with $t$ even.
  \item Let $l=\frac{q^2-1}{s}$ and $m=\frac{q^2-1}{t}.$
  \item Let $g$ be a primitive element of $\F_{q^2}$, $\delta=g^{s}\; \text{and}\; \theta=g^{t}$.
\end{itemize}

Assume $1\leq h\leq \lfloor\frac{s}{2}\rfloor$ and $2\leq r\leq \lfloor\frac{t}{2}\rfloor$. We will construct EAQMDS codes with length $n=lh+mr$. The following lemmas will be used in
our construction.\\

{\bf Lemma 3.1} \quad Assume $a_{0},a_{1}.\cdots,a_{h-1}\in \F_{q}^{*}.$ The following system of equations

\begin{equation}
\left\{
\begin{array}{lll}
u_{0}+u_{1}+\cdots+u_{h-1}=a_{0}\\\\
\sum\limits_{k=0}^{h-1}g^{(2k+1)[(\frac{s-h}{2}+1)l-q-1]}u_{k}=a_{1}\\\\

                 \vdots \\\\
                 \sum\limits^{h-1}_{k=0}g^{(2k+1)[(\frac{s+h}{2}-1)l-q-1]}u_{k}=a_{h-1}\\\\\end{array} \right.\end{equation}

has a solution, denote by $\mathbf{u}=(u_{0},u_{1},\cdots,u_{h-1})\in (\F_{q}^*)^h $.\\

Since the proof is lengthy, we move it to the appendix for the convenience of reading. \\

Let $\mathbf{u}=(u_{0},u_{1},\cdots,u_{h-1})$ satisfy the system of equations (2). Now we divide the vectors $\mathbf{a}$ and $\mathbf{v}$ into two parts and the first part was described as follows.\\
Choose $$\mathbf{a}_{1}=(g,g\delta,\cdots,g\delta^{l-1},g^3,g^3\delta,\cdots,g^3\delta^{l-1},\cdots,g^{2h-1},g^{2h-1}\delta,\cdots,g^{2h-1}\delta^{l-1})$$
and $$\mathbf{v}_{1}=(v_{0},v_{0}\delta,\cdots,v_{0}\delta^{l-1},v_{1},v_{1}\delta,\cdots,v_{1}\delta^{l-1},\cdots,v_{h-1},v_{h-1}\delta,\cdots,v_{h-1}\delta^{l-1})$$
where $v_{k}^{q+1}=u_{k} \quad (0\leq k\leq h-1)$ and $v_{i}\in \F_{q^2}^{*}$.\\

{\bf Lemma 3.2} \quad The inequality $\langle\mathbf{a}_{1}^{qi+j},\mathbf{v}_{1}^{q+1}\rangle\neq 0$ holds if and only if $qi+j+q+1=\mu\cdot l$ , where $\frac{s-h}{2}+1\leq \mu\leq \frac{s+h}{2}-1$.\\

{\bf Proof} \quad When $(i,j)=(0,0)$,
 \begin{align*}
\langle\mathbf{a}_{1}^{0},\mathbf{v}_{1}^{q+1}\rangle &=v_{0}^{q+1}+l(v_{1}^{q+1}+\cdots+v_{h-1}^{q+1})\\
&=u_{0}+l(u_{1}+u_{1}+\cdots+u_{h-1})\neq 0.\end{align*}

When $(i,j)\neq (0,0)$,
 \begin{align*}
\langle\mathbf{a}_{1}^{qi+j},\mathbf{v}_{1}^{q+1}\rangle &=\sum_{k=0}^{h-1}g^{(2k+1)(qi+j)}\cdot v_{k}^{q+1}\sum_{e=0}^{l-1}\delta^{e[(qi+j)+(q+1)]}\\
&=\left\{
\begin{array}{lll}
0, \quad \quad\quad \quad \quad \quad \quad \quad\; \quad \quad \quad \quad \quad \;\text{if}\; l\nmid qi+j+q+1 ,\\\\
l\cdot \sum\limits_{k=0}^{h-1}g^{(2k+1)(qi+j)}u_{k}, \quad \quad \quad \quad\quad  \text{if}\; l\mid qi+j+q+1.
\end{array} \right.\end{align*}

For $0\leq i,j\leq \frac{s+h}{2}\cdot \frac{q-1}{2}-3<q-2$, it is easy to verify $0<qi+j+q+1<q^2-1$.
By Lemma 3.1, there exists $u_{0},u_{1},\cdots,u_{h-1}\in (\F_{q}^*)^h$ satisfying system of equations (2).
Also $l\mid qi+j+q+1$ if and only if $$qi+j+q+1=\left(\frac{s-h}{2}+1\right)\cdot l,\cdots, \left(\frac{s+h}{2}-1\right)\cdot l,$$
since  $0\leq i,j\leq \frac{s+h}{2}\cdot \frac{q+1}{s}-3<q-2\;\text{and}\;0<qi+j+q+1<q^{2}-1$. \\
It follows that $qi+j+q+1=q(\frac{\mu(q+1)}{s}-1)+(q-\frac{\mu(q+1)}{s})$,
which implies $$i=\frac{\mu(q+1)}{s}-2,\; j=q-\frac{\mu(q+1)}{s}-1.$$
Therefore $\frac{s-h}{2}<\mu<\frac{s+h}{2}$. Then $l\mid qi+j$ if and only if $qi+j+q+1=\mu l$, with $\frac{s-h}{2}+1\leq \mu\leq \frac{s+h}{2}-1$.
which completes the proof.$\hfill\qedsymbol$\\

For the second part of coordinates in $\mathbf{a}$ and $\mathbf{v},$ choose

$$ \mathbf{a}_{2}=(1,\theta,\cdots,\theta^{m-1},g^2,g^2\theta,\cdots,g^2\theta^{m-1},\cdots,g^{2r-2},\cdots,g^{2r-2}\theta^{m-1})$$
and
$$ \mathbf{v}_{2}=(1,g^{\frac{t}{2}},\cdots,g^{(m-1)\frac{t}{2}},1,g^{\frac{t}{2}},\cdots,g^{(m-1)\frac{t}{2}},\cdots,1,g^{\frac{t}{2}},\cdots,g^{(m-1)\frac{t}{2}}).$$
Then we have following lemma.\\

{\bf Lemma 3.3} \quad  The identity $$ \langle\mathbf{a}_{2}^{qi+j},\mathbf{v}_{2}^{q+1}\rangle=0$$
holds for all $0\leq i,j \leq \frac{q+1}{2}+\frac{q-1}{t}-2.$ \\

Since the proof is lengthy, we move it to the appendix for the convenience of reading. \\\\

Concatenating $\mathbf{a}_{1}$ and $\mathbf{a}_{2}$, $\mathbf{v}_{1}$ and $\mathbf{v}_{2}$, we obtain $\mathbf{a}=(\mathbf{a}_{1}|\mathbf{a}_{2})$, $\mathbf{v}=(\mathbf{v}_{1}|\mathbf{v}_{2})$. Then the inner product $\langle\mathbf{a},\mathbf{v}\rangle=\langle\mathbf{a}_{1},\mathbf{v}_{1}\rangle+\langle\mathbf{a}_{2},\mathbf{v}_{2}\rangle.$
By Lemma 3.3,
  $$ \langle\mathbf{a}^{qi+j},\mathbf{v}^{q+1}\rangle= \langle\mathbf{a}_{1}^{qi+j},\mathbf{v}_{1}^{q+1}\rangle .$$
Notations as in Lemma 3.2, it follows that the inequality $\langle\mathbf{a}_{1}^{qi+j},\mathbf{v}_{1}^{q+1}\rangle\neq 0$  holds if and only if  $qi+j+q+1=\mu\cdot l$ for some $l$. \\

{\bf Theorem 3.1}\quad Put $n=lh+mr,c=h-1.$ There exists an EAQMDS code with parameters $$[[n,n-2d+c,d+1;c]]_{q}$$
where $1\leq d\leq min\left\{\frac{s+h}{2s}\cdot (q+1)-2, \frac{q+1}{2}+\frac{q-1}{t}-1\right\}.$\\

{\bf Proof}\quad Let $G$ be a generator matrix of $\mathbf{GRS}_{k}(\mathbf{a},\mathbf{v})$ in (1).  Therefore, its Hermitian dual code $\mathbf{GRS}_{n-k}(\mathbf{a}^q,\mathbf{v^\prime})$ has parameters $[n,n-d,d+1]$ with parity check matrix $G^q$.
It suffices to prove rank$(GG^{\dag})=c$ with $G^{\dag}$ being conjugate transpose of $G$. Denote $\langle\mathbf{a}^{qi+j},\mathbf{v}^{q+1}\rangle$ by $a_{i,j}$. A calculation shows
$$GG^{\dag}=\left(
              \begin{array}{cccc}
                a_{0,0} & a_{1,0} & \cdots & a_{d-1.0} \\
                \vdots & \vdots & \vdots & \vdots \\
                a_{0,d-1} & a_{1,d-1} & \cdots & a_{d-1,d-1} \\
              \end{array}
            \right)
.$$

For any $1\leq d\leq \text{min}\{\frac{s+h}{2s}\cdot (q+1)-2, \frac{q+1}{2}+\frac{q-1}{t}-1\}$. By Lemmas 3.2 and 3.3, the matrix $(GG^\dag)$ has
exactly $h-1$ nonzero entries which are in different rows and columns. Therefore, it is easy to verify rank$(GG^\dag)=c$. By Theorem 2.1, there exists an EAQMDS code with parameters $[[n,n-2d+c,d+1;c]]_{q}$. This completes the proof.$\hfill\qedsymbol$\\

{\bf Remark 3.1}\quad Compared to the known results listed in table 1, the length of our codes could be the sum of the factor of $q^2-1$ which hasn't been reported before.
For given $q$, the range of length is $\lceil\frac{t+2s}{st}(q^2-1)\rceil\leq n\leq q^2-1$.  When $h$ approaches to $\frac{s}{2}$ and $t=4$, the minimal distance $d$ can reach $\frac{3}{4}q$.\\\\

{\bf Example 3.\,1}\quad We list some new parameters of EAQMDS codes of Theorem 3.1 in Table 2.\\\\

\begin{center}
\begin{longtable}{cccccc}  
\caption{Some new parameters of EAQMDS codes } \\  \hline

Parameters & $s$ & $h$ &$r$ &$t$ &$d$\\ \hline
$[[640,641-2d,d+1;1]]_{29}$  & 6  & 2  & 3 & 7 &$1\leq d\leq 18$ \\ \hline

$[[300,301-2d,d+1;1]]_{19}$  & 4  & 2  & 2 & 6 &$1\leq k\leq 12$   \\  \hline

$[[696,697-2d,d+1;1]]_{29}$  & 5  & 2 & 3 & 7 &$1\leq k\leq 18$ \\   \hline

$[[128,130-2d,d+1;2]]_{13}$  & 7 & 3 & 2 & 6  &$1\leq k\leq 8$  \\   \hline

\end{longtable}
 \end{center}

\section{Second Construction}

If there are no specific statements, the following notations are fixed throughout this section.
\begin{itemize}
  \item Let $2s+1\mid q+1$ and $t=\frac{q^2-1}{2s+1}$.
  \item Let $c=2e+1$ and $0\leq e \leq s-1.$
  \item Let $g\in \F_{q^2}$ be a primitive element and $\gamma=g^{2s+1}$.
\end{itemize}
{\bf Lemma 4.1} \quad The following system of equations

\begin{equation}\label{4.1eq}
\left\{
\begin{array}{lll}

u_{1}+u_{2}+\cdots+u_{c}=a_0\\\\

g^{(s-e+1)t}u_{1}+g^{2(s-e+1)t}u_{2}+\cdots+g^{c(s-e+1)t}u_{r}=a_1\\\\
g^{(s-e+2)t}u_{1}+g^{2(s-e+2)t}u_{2}+\cdots+g^{c(s-e+2)t}u_{r}=a_2\\\\
 \vdots \\\\

g^{(s+e)t}u_{1}+g^{2(s+e)t}u_{2}+\cdots+g^{c(s+e)t}u_{r}=a_{c-1}\\\\
\end{array} \right.
\end{equation}
has a solution $\mathbf{u^{\prime}}=(u_{1},u_{2},\cdots,u_{c})\in (\F_{q}^{*})^{c}.$\\

{\bf Proof} \quad Let
$$D=\left(
  \begin{array}{cccc}
    1 & 1 & \cdots & 1 \\
    g^{(s-e+1)t} & g^{2(s-e+1)t} & \cdots & g^{c(s-e+1)t} \\
    \vdots & \vdots & \ddots & \vdots \\
    g^{(s+e)t} &  g^{2(s+e)t} & \cdots & g^{c(s+e)t} \\
  \end{array}
\right).$$
Then the system of the equations (\ref{4.1eq}) is equivalent to
$$ D\mathbf{u^{\prime}}^{T}=\mathbf{\xi}^{T},$$
where $\xi=(a_{0},a_{1},\cdots,a_{c-1}).$

By Lemma 3.1, we know that the matrix is invertible.
Then there exists only one solution $\mathbf{u^{\prime}}=(u_{1},u_{2},\cdots,u_{c})\in (\F_{q}^{*})^{c}$.  This completes the proof.    $\hfill\qedsymbol$\\

It is easy to verify that $ \gamma =g^{2s+1}$
is a primitive $t$-th root of unity and  $$g^{i_{1}}\gamma^{j_{1}}\neq g^{i_{2}}\gamma^{j_{2}}$$  where $1\leq i_{1}\neq i_{2}\leq c$ and $0\leq j_{1}\neq j_{2}\leq t-1.$\\

\noindent Put
$$ \mathbf{a_{3}}=(0,g,g\gamma,\cdots,g\gamma^{t-1},g^{2},g^{2}\gamma,\cdots,g^{2}\gamma^{t-1},\cdots,                                   g^{c},g^{c}\gamma,\cdots,g^{c}\gamma^{t-1})\in \F_{q^2}^{n},$$

$$ \mathbf{v_{3}}=(b_{0},b_{1},\cdots,b_{1},\cdots,b_{c},\cdots,b_{c})$$
where $b_{i}^{q+1}=u_{i} \quad (1\leq i\leq c)$ and $u_{i}\in \F_{q}^{*}$.\\

{\bf Lemma 4.2} \quad  The identity $\langle \mathbf{a_{3}}^{qi+j},\mathbf{v_{3}}^{q+1}\rangle\neq 0$ if and only if
$qi+j=(s-e+1)m,(s-e+2)m,\cdots,(s+e)m.$\\

{\bf Proof} \quad Note that $2s+1\,\mid q+1$ and $1 \leq c \leq t.$  By Lemma 4.1, there exists a solution of the system of equations $(\ref{4.1eq})$, say $\mathbf{u^{\prime}}=(u_{1},u_{2},\cdots,u_{c})\in (\F_{q}^{*})^{c}$.\\

 For $(i.j)=(0,0),$ let $b_{i}\in \F_{q^2}$ such that $b_{i}^{q+1}=u_{i}(i=1,2,\cdots,c)$ and $b_{0}\in \F_{q^2}^{*}$ satisfies $b_{0}^{q+1}\neq -ta_{1}$.
 Then
\begin{align*}
\langle\mathbf{a}_{3}^{0},\mathbf{v}_{3}^{q+1}\rangle &=b_{0}^{q+1}+t(b_{1}^{q+1}+b_{2}^{q+1}+\cdots +b_{c}^{q+1})\\
&=b_{0}^{q+1}+t(u_{1}+u_{2}+\cdots+u_{c})\\
&=b_{0}^{q+1}+ta_{1}\neq 0.\end{align*}

For any $(i.j)\neq (0,0),$ a straightforward calculation shows

$$ \langle\mathbf{a_{3}}^{qi+j},\mathbf{v_{3}}^{q+1}\rangle=\sum_{l=1}^{c}{g^{l(qi+j)}}b_{l}^{q+1}\sum_{e=0}^{t-1}{\gamma^{e(qi+j)}},$$
which implies
$$\langle\mathbf{a_{3}}^{qi+j},\mathbf{v_{3}}^{q+1}\rangle=\left\{
\begin{array}{lll}
0, \quad \quad \quad \quad \quad \quad \quad \quad \quad \quad \quad  \text{if}\; t\nmid qi+j, \\\\

t\sum\limits_{l=1}^{c}g^{l(qi+j)}u_{l}, \,\,\;\quad \quad \quad \quad\quad  \text{if}\; t\mid qi+j \quad \text{and} \quad(i,j)\neq (0,0).
\end{array} \right.$$

Since $1\leq k \leq (s+1+e)\frac{q+1}{2s+1}-1,0\leq i,j\leq k-1.$ The next proof is similar to Lemma 3.2.\\

Since $t\mid qi+j$ if and only if $$qi+j=(s-e+1)t,(s-e+2)t,\cdots,(s+e)t.$$
It is easy to see that $k\leq q+1$. Hence for any $0\leq i,j\leq k-1$, we have $qi+j<(q+1)k\leq q^{2}-1$.
Assume $(i,j)\neq (0,0)$. If $qi+j=lt=l\frac{q^2-1}{2s+1}$, then $0<l<2s+1$.
Note that $qi+j=l\frac{q^{2}-1}{2s+1}=q(\frac{l(q+1)}{2s+1}-1)+(q-\frac{l(q+1)}{2s+1}),$
which implies $i=\frac{l(q+1)}{2s+1}-1,j=q-\frac{l(q+1)}{2s+1}$. It suffices to prove the following two cases.

{\bf Case 1:}\quad If $l\geq s+1+e$, then $i=\frac{l(q+1)}{2s+1}-1\geq (s+e+1)\frac{q+1}{2s+1}-1\geq k$, which contradicts to the assumption that $i\leq k-1$.\\

{\bf Case 2:} \quad If $l<s-e$, then $j=q-\frac{l(q+1)}{2s+1}\geq (s+1+e)\frac{q+1}{2s+1}-1\geq k$ , which contradicts to the assumption that $j\leq k-1$. Thus $s-e+1\leq l\leq s+e$.

Therefore, when $qi+j=(s-e+1)m,(s-e+2)m,\cdots,(s+e)m,$  we have $$\langle\mathbf{a_{3}}^{qi+j},\mathbf{v_{3}}^{q+1}\rangle\neq 0.$$
It follows that
$\langle\mathbf{a_{3}}^{qi+j},\mathbf{v_{3}}^{q+1}\rangle\neq 0$ if and only if $qi+j=(s-e+1)m,(s-e+2)m,\cdots,(s+e)m$. This completes the proof.$\hfill\qedsymbol$\\

{\bf Theorem 4.1} \quad (1) Let $n=1+(2e+1)\frac{q^2-1}{2s+1}$, for $c=2e$ and $1\leq k \leq (s+1+e)\frac{q+1}{2s+1}-1.$ There exists EAQMDS code with parameters
$$\left[\left[1+(2e+1)\frac{q^2-1}{2s+1},1+(2e+1)\frac{q^2-1}{2s+1}-2k+c, k+1; c\right]\right]_{q}$$ where $2s+1 \mid q+1$ and $0\leq e \leq s-1.$

(2)  Let $n=1+(2e+2)\frac{q^2-1}{2s}$, for $c=2e+1$ and $1\leq k \leq (s+1+e)\frac{q+1}{2s}-1.$ There exists EAQMDS code with parameters
$$\left[\left[1+(2e+2)\frac{q^2-1}{2s}, 1+(2e+2)\frac{q^2-1}{2s}-2k+c, k+1; c\right]\right]_{q}$$ where $2s \mid q+1$ and $0\leq e \leq s-2.$

(3)  Let $n=1+(2e+1)\frac{q^2-1}{2s}$, for $c=2e$ and $1\leq k \leq (s+e)\frac{q+1}{2s}-2.$ There exists EAQMDS code with parameters
$$\left[\left[1+(2e+1)\frac{q^2-1}{2s}, 1+(2e+1)\frac{q^2-1}{2s}-2k+c, k+1; c\right]\right]_{q}$$ where $2s \mid q+1$ and $0\leq e \leq s-1.$\\

{\bf Proof}\quad For simplicity, we only prove the case (1), the other two cases are similar and we omit the details.

It is easy to verify that the entries of $\mathbf{a}_{3}$ are distinct in $\F_{q^2}$. The code $\mathbf{GRS}_{k}(\mathbf{a}_{3},\mathbf{v}_{3})$ has generator matrix

$$G_{1}=\left(
      \begin{array}{cccccccc}
        b_{0} & b_{1} & \cdots & b_{1} & \cdots & b_{c} & \cdots & b_{c}  \\
        0 & b_{1}g & \cdots & b_{1}g\gamma^{t-1} & \cdots & b_{c}g^{c} & \cdots & b_{c}g^{c}\gamma^{t-1}  \\
        0 &  b_{1}g^2 & \cdots &  b_{1}(g\gamma^{t-1})^2 & \cdots & b_{c}g^{2c} & \cdots & b_{c}(g^{c}\gamma^{t-1})^2  \\
        \vdots & \vdots  & \vdots  & \vdots & \vdots  & \vdots & \vdots & \vdots  \\
        0 & b_{1}g^{k-1} & \cdots & b_{1}(g\gamma^{t-1})^{k-1} & \cdots  & b_{c}g^{c(k-1)} & \cdots & b_{c}(g^{c}\gamma^{t-1})^{k-1}  \\
      \end{array}
    \right).$$
There exists a $\mathbf{GRS}_{n-k}(\mathbf{a}^q_{3},\mathbf{v}^\prime_{3})$ code which is the dual of $\mathbf{GRS}_{k}(\mathbf{a}_{3},\mathbf{v}_{3})$ code.
And the code $\mathbf{GRS}_{n-k}(\mathbf{a}_{3},\mathbf{v}^\prime_{3})$  has parameters $[n,n-k,k+1]$ with parity check matrix $G^q_{1}$ ($G^q$ means raising every entry of $G$ to $q$-th power). Denote $\langle\mathbf{a}_{3}^{qi+j},\mathbf{v}_{3}^{q+1}\rangle$ by $a_{i,j}$. It is easy to calculate

$$GG^{\dag}=\left(
              \begin{array}{cccc}
                a_{0,0} & a_{1,0} & \cdots & a_{k-1,0} \\
                 \vdots& \vdots & \vdots & \vdots \\
                a_{0,k-1} & a_{1,k-1} & \cdots & a_{k-1,k-1} \\
              \end{array}
            \right).$$

For $1\leq k \leq (s+1+e)\frac{q+1}{2s+1}-1$. By Lemma 4.2, it is easy to verify rank$(GG^{\dag})=c$. According to Theorem 2.1, there exists EAQMDS code
with parameters $$[[n, n-2k+c, k+1; c]]_{q}.$$
This completes the proof. $\hfill\qedsymbol$\\

{\bf Remark 4.1}\quad In our constructions, the length of our codes are not a factor of $q^2\pm 1$. The parameters of our results are very flexible while
the required number of maximally entangled states of many known EAQMDS codes
reported in the literature (see Table 1) is a fixed number. For given $q$, the length of our codes can reach $(q-3)(q+1)$. The range of minimal distance $d$ is $2\leq d\leq \lfloor\frac{2s-1}{2s}(q+1)\rfloor$.  Therefore, new EAQMDS codes can be obtained by Theorems 3.1 and 4.1. Some of them are listed in Table 3.\\

{\bf Example 4.\,1}\quad We list some new parameters of EAQMDS codes of Theorem 4.1 in Table 3.\\\\

\begin{center}
\begin{longtable}{ccccc}  
\caption{Some new parameters of EAQMDS codes } \\  \hline

Parameters & $s$ & $e$ &$k$ & Case of Theorem 4.1\\ \hline
$[[121,125-2k,k+1;4]]_{13}$  & 3  & 2  &  $1\leq k\leq 11$ & Theorem 4.1(1) \\ \hline

$[[225,231-2k,k+1;6]]_{17}$  & 4  & 3 & $1\leq k\leq 15$  & Theorem 4.1(1)  \\  \hline

$[[145,156-2k,k+1;12]]_{13}$  & 7  & 5 &$1\leq k\leq 12$ & Theorem 4.1(2)\\   \hline

$[[157,169-2k,k+1;12]]_{13}$  & 7 & 6 &$1\leq k\leq 11$  & Theorem 4.1(3)\\   \hline

\end{longtable}
 \end{center}

\section{Appendix}
{\bf Proof of lemma 3.1} \quad Denote by $\xi=(a_{0},a_{1},\cdots,a_{h-1})$ and $d=\frac{s-h}{2}+1 $.
The system of equations (2) can be expressed in the matrix form $$A\cdot \mathbf{u}^{T} =(a_{0},a_{1},\cdots,a_{h-1})^{T}=\xi^{T},$$
where $$A=\left(
            \begin{array}{cccc}
              1 & 1 & \cdots & 1 \\
              g^{dl-q-1} & g^{3(dl-q-1)} & \cdots & g^{(2h-1)(dl-q-1)} \\
              \vdots & \vdots & \vdots & \vdots \\
              g^{(d+h-2)l-q-1} & g^{3[(d+h-2)l-q-1]} & \cdots & g^{(2h-1)[(d+h-2)l-q-1]} \\
            \end{array}
          \right)
.$$
Let $x=g^{dl-q-1}$ and $$A^{T}=\left(
            \begin{array}{cccc}
              1 & x & \cdots & xg^{(h-2)l} \\
              1 & x^{3} & \cdots & (xg^{(h-2)l})^{3} \\
              \vdots & \vdots & \vdots & \vdots \\
              1 & x^{2h-1} & \cdots & (xg^{(h-2)l})^{2h-1} \\
            \end{array}
          \right)
.$$\\
A routine calculation shows
$$\text{det}(A^T)=\frac{1}{x\cdot(xg^l)\cdot\cdots\cdot(xg^{(h-2)l})}\left|
                                                                  \begin{array}{cccc}
                                                                    1 & 1 & \cdots & 1 \\
                                                                    1 & x^{2} & \cdots & (xg^{(h-2)l})^{2} \\
                                                                    \vdots & \vdots & \vdots& \vdots \\
                                                                    1 & x^{2h-2} & \cdots & (xg^{(h-2)l})^{2h-2} \\
                                                                  \end{array}
                                                                \right|
.$$\\

It is obvious that $\text{det}(A)\neq 0$ and $\mathbf{u}=(u_{0},u_{1},\cdots,u_{h-1})\in (\F_{q^2})^h $.\\

Next we will show that (2) has a solution $\mathbf{u}=(u_{0},u_{1},\cdots,u_{h-1})\in (\F_{q^*})^h $.
Let
 $$\mathbf{\beta_{0}}=(1,1,\cdots,1),$$
  $$\mathbf{\beta_{i}}=\left(g^{(d+i-1)l-q-1},g^{3[(d+i-1)l-q-1]},\cdots,g^{(2h-1)[(d+i-1)l-q-1]}\right) \text{for}\; i=1,2,\cdots,h-1.$$

It is easy to verify $\beta_{i}^q=\beta_{h-i}$ for $i=1,2,\cdots,h-1.$ Assume $\zeta=g^{\frac{q+1}{2}}$, then $\zeta^{q}=-\zeta.$
Raising all the equations of (2) to $q$-th power leads to
\begin{equation}\label{conj}
\left\{
\begin{array}{lll}
\beta_{0}\mathbf{u}=a_{0}\\\\
(\beta_{1}+\beta_{h-1})\mathbf{u}=2a_{1}\\\\

                 \vdots \\\\
                 (\beta_{\frac{h-1}{2}}+\beta_{\frac{h+1}{2}})\mathbf{u}=2a_{\frac{h-1}{2}}\\\\
                  \frac{\beta_{1}-\beta_{h-1}}{\zeta}\mathbf{u}=0\\\\
                 \vdots \\\\
                 \frac{\beta_{\frac{h-1}{2}}-\beta_{\frac{h+1}{2}}}{\zeta}\mathbf{u}=0.\end{array} \right.\end{equation}
Denote by $\alpha=\left(a_{0},2a_{1},\cdots,2a_{\frac{h-1}{2}},0,\cdots,0\right)^T$.  The above equations (\ref{conj}) can be reformulated as $B\mathbf{u}=\alpha$. It is easy to verify that all the entries of $B$ are in $\F_{q}$.
Suppose $$S=\left\{\mathbf{u}\in \F_{q}^{n}|B\mathbf{u}=\alpha,\;a_{i}\in\F^*_q, 1\leq i\leq h-1\right\}.$$
Then $|S|=(q-1)^{\frac{h+1}{2}}$ since $B$ is invertible. For $0\leq i\leq h-1$, define
$$ S_{i}=\left\{\mathbf{u}\in \F_{q}^{n}|u_{i}=0,\;B\mathbf{u}=\alpha,\;a_{j}\in\F^*_q\;\text{for all}\,0\leq j\leq \frac{h-1}{2}\; \right\}\;.$$
Then the set of solutions to (\ref{conj}), denote by $V$, is
 $$V=\overline{ S_{1}\bigcup S_{2}\bigcup \cdots \bigcup S_{\frac{h+1}{2}}},$$
where $\overline{T}=S-T$ for any subset $T\subseteq S.$
By Inclusion-Exclusion principle,
$$|V|=|S|-\sum\limits^{h-1}_{i=0}|S_{i}|+r$$
for some $r\geq 0.$
It follows that$$|V|=(q-1)^{\frac{h+1}{2}}-\binom{h}{1}(q-1)^{\frac{h-1}{2}}+r.$$
 Obviously $|V|>0$.
We only consider the case $h$ is odd.  The case $h$ being even is similar and we omit the details.
Therefore, $u_{i}\in\F_{q}^{*}(i=0,1,\cdots,h-1).$

This completes the proof. $\hfill\qedsymbol$\\

\noindent{\bf Proof\; of \;Lemma\; 3.3}\quad  We can calculate directly,
\begin{align*}
\langle\overrightarrow{a}_{2}^{qi+j},\overrightarrow{v}_{2}^{q+1}\rangle &=\sum_{k=0}^{r-1}\sum_{\nu=0}^{m-1}(g^{k}\theta^{\nu})^{qi+j}\cdot \theta^{\nu\cdot \frac{q+1}{2}}\\
&=\sum_{k=0}^{r-1}g^{k(qi+j)}\sum_{\nu=0}^{m-1}\theta^{\nu(qi+j+\frac{q+1}{2})}.\end{align*}
It suffices to prove the identity
\begin{align*}\sum_{v=0}^{m-1}\theta^{v(qi+j+ \frac{q+1}{2})}=0\end{align*}
holds for all $ 0\leq i,j\leq \frac{q+1}{2}+ \frac{q-1}{t}-2 ,\;t \geq 2.$
It is easy to check that the identity holds if and only if $m\nmid qi+j+\frac{q+1}{2}$. On the contrary, assume that $m\mid qi+j+\frac{q+1}{2}$. Let
 \begin{equation}\label{eq3}
    qi+j+\frac{q+1}{2}=\mu\cdot m=q\cdot \frac{\mu(q-1)}{t}+\frac{\mu(q-1)}{t}
 \end{equation}
where $\mu$ is an integer. Since $t\geq 2$, we have $qi+j+\frac{q+1}{2}<q^{2}-1$, which implies $0<\mu<t$. It suffices to prove the following two cases.\\

{\bf Case 1:}\quad If $j+\frac{q+1}{2}\leq q-1$, comparing remainder and quotient of module $q$ on both sides of (\ref{eq3}), we can deduce $i=j+\frac{q+1}{2}=\mu\cdot \frac{q-1}{t}$. Since $t$ is even, then $\frac{q-1}{t}\mid \frac{q-1}{2}$. From $\frac{q-1}{t}\mid j+1+\frac{q-1}{2}$, we can deduce that $\frac{q-1}{t}\mid j+1$. Since $j+1\geq 1$, then $j+1\geq \frac{q-1}{t}$.
Therefore, $i=j+\frac{q+1}{2}\geq \frac{q+1}{2}+\frac{q-1}{t}-1$, which is a contradiction.\\

{\bf Case 2:}\quad If $j+\frac{q+1}{2}\geq q$, it takes $$qi+j+\frac{q-1}{2}=q(i+1)+\left(j-\frac{q-1}{2}\right)=q\cdot \frac{\mu(q-1)}{t}+\frac{\mu(q-1)}{t}.$$
In a similar way, $j-\frac{q-1}{2}=i+1=\mu\cdot \frac{q-1}{t}$ which implies $\frac{q-1}{t}\mid i+1$. Since $i+1\geq 1$, then $i+1\geq \frac{q-1}{t}$. Therefore, $j=i+1+\frac{q+1}{2}\geq \frac{q+1}{2}+\frac{q-1}{t}-1$, which is a contradiction.

As a result, $m\nmid qi+j+\frac{q+1}{2}$ which yields \begin{align*}\sum_{v=0}^{m-1}\theta^{v(qi+j+ \frac{q+1}{2})}=0\end{align*}
 for all $ 0\leq i,j\leq \frac{q+1}{2}+ \frac{q-1}{t}-2 .$ This completes the proof.$\hfill\qedsymbol$\\

\section{Conclusion}

In this paper, we present some new constructions of EAQMDS codes from generalized Reed-Solomon codes. Comparing to the known results, the parameters of our
EAQMDS codes are flexible. It is expected that more new types of EAQMDS codes could be constructed if the number of shared pairs can be explicitly determined.

\end{document}